\documentstyle[12pt,epsfig]{article} 
\title{
\rightline{\small UL-NTZ 03/96}
Quantum effects in radiation on short bunches}

\author{R.~Engel$^1$, A.~Schiller$^1$ and  V.G.~Serbo$^{1,2}$  \\
{\it $^1$ Institut f\"ur Theoretische Physik, Universit\"at
Leipzig, F.R. Germany}  \\
{\it $^2$ Novosibirsk State University, Novosibirsk, Russia}}

\date{January 31, 1996}

\begin{document}

\maketitle
\begin{abstract}
The radiation caused by particles of one bunch in
the collective electromagnetic field of the short oncoming bunch 
is studied.
Quantum effects are calculated for the spectrum of radiated photons.
Using this spectrum, the dependence of the relative
energy loss $\delta$ on a quantum parameter $\kappa$
is dicussed. It is shown that the behaviour of
$\delta$ changes considerably with the increase of that parameter.
In the classical regime ($\kappa \ll 1$) the energy
loss is proportional to the incoming particle energy,
while in the extreme quantum regime
($\kappa \gg 1$) the energy loss becomes a constant. 
The coherent $e^+e^-$ pair production for $\gamma e$ colliders as
cross-channel to CBS is considered.
\end{abstract}

\section{Introduction}

There are two types of coherent radiation at colliders ---
beamstrahlung (BS) and coherent bremsstrahlung (CBS).
Beamstrahlung takes place on colliders with long bunches when the
deflection angle of a radiating particle $\theta_d$ with Lorentz
factor $\gamma_e$ is much larger than the typical radiation angle
$\theta_r \sim 1/ \gamma_e$, it occurs mainly at   $e^+e^-$
linear colliders. CBS occurs at most of the existing colliders
having short bunches ($\theta_d \ll \theta_r$).  The different
types of coherent radiation can be characterized by the parameter
$\eta$ which is related to these angles via\footnote{Restricting
ourselves to $e^+e^-$ colliders, we consider, for definiteness,
the photon radiation by electrons moving through a positron
bunch. We denote by $N_e$ and $N_p$ the numbers of particles in
the electron and positron bunches. $\sigma_z$ is the
longitudinal, $\sigma_x$ and $\sigma_y$ are the horizontal and
vertical transverse sizes of the positron bunch,
$\gamma_e=E_e/m_ec^2$ is the electron Lorentz factor and
$r_e=e^2/m_e c^2$ is the electron classical radius.}
\begin{equation}
{\theta_d \over \theta_r} \sim \eta = { N_p r_e \over \sigma_x
+\sigma_y} \ .
\label{0}
\end{equation}

Classical and quantum regimes for BS ($\eta \gg 1$)
are discussed in a number of papers (see, for example, the review
\cite{ChenR}).  For CBS ($\eta \ll 1$), only the classical regime has been 
considered up to now \cite{Bassetti}-\cite{ESS}. 
In our previous paper \cite{ESS}
we have presented a simple method for calculating CBS based on a
developed equivalent photon approximation for  coherent
processes. Here we apply this method to study the quantum
effects for CBS.  They are characterized by the quantum parameter
$\kappa$ which is   defined by the ratio
\begin{equation}
\kappa= {E_c \over E_e}\ ,
\label{01}
\end{equation}
where the critical energy $E_c$ for CBS is given by the expression
\begin{equation}
E_c={4\gamma_e^2 \hbar c \over \sigma_z} \ .
\label{02}
\end{equation}
The classical regime is characterized by $\kappa \ll 1$, the
extreme quantum limit corresponds  to $\kappa \gg 1$.

In Sect.~2 the spectrum of CBS photons is calculated as function
of the photon energy $E_\gamma$ and the quantum parameter
$\kappa$. 
Additionally
both quantum corrections to the classical case as well as
corrections to the extreme quantum case are presented.  Using the
results of Sec.~2, the relative energy loss is discussed in
Sect.~3.  In the next section we discuss quantitatively a
possibility to reduce the beamstrahlung energy loss using CBS
bunchlets. Finally, we present in Sect.~5 a cross-channel to CBS
-- the coherent pair production at $\gamma e$ colliders. The
number of produced $e^+e^-$ pairs and the energy spectrum is
calculated.  Our main results are summarized in the Conclusions.

\section{Spectrum of CBS}

The energy spectrum of CBS photons is given in detail in
\cite{ESS}.  Here, we summarize the results which are important
for the following discussion.  The number of CBS photons for a
single collision of the bunches is
\begin{equation}
dN_\gamma = N_0\;\Phi ({E_\gamma / E_e} , \kappa )\; {dE_\gamma
\over E_\gamma} \ .
\label{2}
\end{equation}
The  dimensionless constant $N_0$ is defined as
\begin{equation}
N_0= {8\over 3}\;\alpha\;r_e^2\;J(0) \ ,
\label{1}
\end{equation}
the function $J(\omega)$ can be found in \cite{ESS}.  For
identical Gaussian beams the constant  $N_0$ is well approximated
by (see Appendix)
\begin{equation}
N_0\approx  0.5 \ \alpha \ N_e \ \eta^2   \ .
\label{111}
\end{equation}

The spectral function
\begin{equation}
\Phi(y, \kappa)={3\over 2} \int_{0}^{\infty}{dz\over (1+z)^2}\;
\left[ {1+z^2 \over (1+z)^2 }  ( 1-y) +{1 \over 2}\,y^2 \right] \
{J(\omega )\over J(0)}
\label{3}
\end{equation}
with
\begin{equation}
y={E_\gamma \over E_e}
\label{331}
\end{equation}
is normalized by the condition
\begin{equation}
\Phi(0, \kappa)=1 \ .
\label{4}
\end{equation}
The integration variable $z$ is related to the polar angle
$\theta$ of CBS photons $z=(\gamma_e \theta)^2$.
The energy  $\hbar \omega$ appearing in $J(\omega)$ is defined by
\begin{equation}
\hbar \omega = (1 +z) \  { E_e \over 4 \gamma_e^2} \ { y \over 1-y }  \ .
\label{41}
\end{equation}

For the practically important case of   Gaussian beams one has
\begin{equation}
{J(\omega )\over J(0)} = \exp{\left[- \left( { \omega  \sigma_z
\over c } \right)^2 \right]} = \exp{\left[ - \left( {  1+z
\over \kappa} \ { y \over  1-y }\right)^2 \right]} \ .
\label{4a}
\end{equation}
In this case the spectral function simplifies to
\begin{equation}
\Phi(y, \kappa) = (1-y) \ \Phi_1(u) + {3\over 4}\, y^2 \
\Phi_2(u) \ , \ \  \  u = {1  \over \kappa} \ { y \over 1-y }
\label{44}
\end{equation}
where the functions $\Phi_i (u)$ are defined as
\begin{equation}
\Phi_1(u)={3\over 2}
\int_{0}^{\infty}\; {1+z^2\over (1+z)^4} \mbox{exp} \left[ -
(1+z)^2 u^2 \right] \ dz \ ,
\label{5}
\end{equation}
\begin{equation}
\Phi_2(u)= \int_{0}^{\infty}\;
{1 \over (1+z)^2} \mbox{exp}
\left[ - (1+z)^2 u^2 \right] \  dz \ .
\label{6}
\end{equation}
The  expansions of $\Phi_i (u)$ for small and large $u$ are given
by
\begin{equation}
\Phi_1(u)  = \left\{
\begin{array}
     {r@{\quad  \quad}l}  1 - {3 \over 2} \sqrt{\pi}  \ u
     \ ,  & u \ll 1\\
     {3 \over 4 u^2} \left(1 -{5 \over 2u^2} +{37\over 4u^4
} + \dots \right) \ \exp{(-u^2)} \ ,  & u \gg 1
\end{array}
\right.
\label{7}
\end{equation}
and
\begin{equation}
\Phi_2(u)  = \left\{
\begin{array}
     {r@{\quad  \quad}l}  1 -   \sqrt{\pi}  \ u            \ ,  & u \ll 1\\
     {1 \over 2 u^2} \left(1 -{3 \over 2u^2} +{15\over 4u^4
} + \dots \right) \ \exp{(-u^2)} \ ,  & u \gg 1 \ .
\end{array}
\right.
\label{8}
\end{equation}

\begin{figure}[htb]
\begin{center}
\epsfig{file=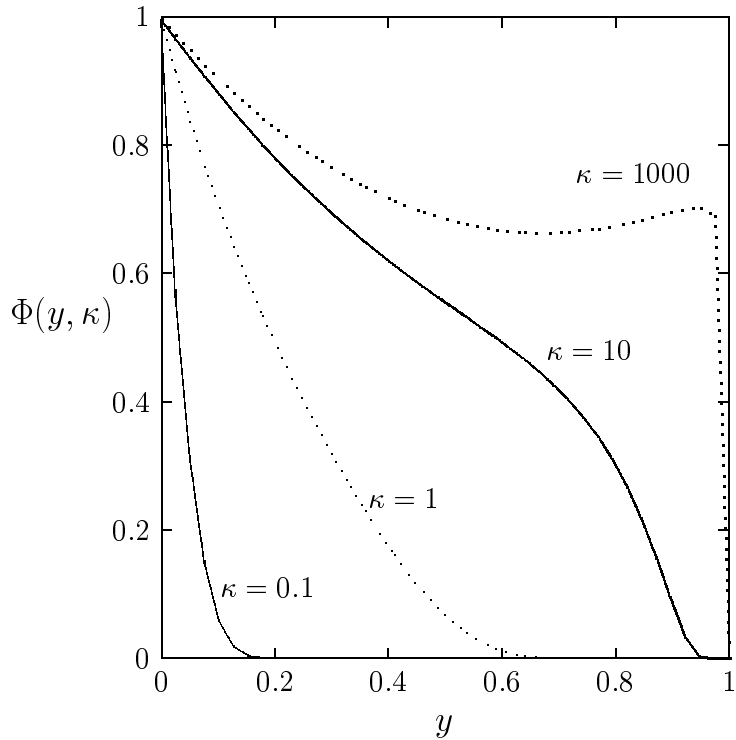,width=10cm}
\caption{Spectral function  $\Phi(y,\kappa)$ (see
(\protect\ref{44})) as function of the energy fraction
$y=E_\gamma/E_e$ for different quantum parameters $\kappa$
\label{qcbs-fig1}}
\end{center}
\end{figure}

To see the transition from the classical to the quantum regime we
show in Fig.~\ref{qcbs-fig1}
the spectrum of CBS photons for different values of the parameter
$\kappa$.  In the classical case ($\kappa \ll 1 $) the spectral
function depends on the ratio $ y / \kappa$ only
\begin{equation}
\Phi(y, \kappa) =  \Phi_1(y/\kappa)=  \Phi_1(E_\gamma /E_c) \ ,
\label{9}
\end{equation}
and photons with energies larger than $E_c$ practically do not
contribute to the distribution. With increasing $\kappa$ the
fraction of high  energy photons increases.

In the extreme quantum regime the quantity $\omega \sigma_z/c$ 
(see (\ref{4a})) is much smaller than 1 for almost all values of 
$E_\gamma$, excluding the region of $E_\gamma$ close to $E_e$. 
Therefore, the ratio $J(\omega)/J(0)$ can be replaced by one and the 
spectral function (at $1-y \gg 1/ \kappa$) simplifies to
\begin{equation}
\Phi (y, \kappa)=1- y +{3\over 4}\, y^2 \ .
\label{10}
\end{equation}
In other words, the positron bunch acts as a particle without
inner structure, and the spectral function does not depend on
the details of the bunch densities.  It is clear that this
function coincides with the spectral function of electron
radiation on a point--like particle. Remind that the standard
cross section for the electron radiation in the incoherent
process $e^-e^+\to e^-e^+ \gamma$ is (see \cite{BLP}, \S 97)
\begin{equation}
d\sigma ={16\over 3}\alpha r_e^2 \ \left(1- y +{3\over 4}\, y^2
\right) \ {dy \over y} \left( \ln{{4E_pE_e(1-y)\over m_e^2 c^4y}
- {1\over 2}}\right)\ .
\label{gam-rad}
\end{equation}
Eq.~(\ref{gam-rad}) contains the same spectral function as (\ref{10}).

This universal dependence on $y$ is violated only near the
kinematical boundary $y \approx 1$ where
\begin{equation}
{1 -y} < {1\over \kappa}\ll 1 \ .
\label{11a}
\end{equation}
For these photon energies the spectrum decreases very sharply
(see Fig.~\ref{qcbs-fig1})  and the shape depends for Gaussian
beams only on the longitudinal size of the positron bunch (in
general on   its density)
\begin{equation}
\Phi(y, \kappa)= {3 \over 8 u^2} \ \exp{(-u^2)} \ , \ \ \ u \gg 1 \ .
 \label{11}
\end{equation}

\section{Relative energy loss}

The knowledge and the control of energy losses is one of the
important collider physics problems. 

We define the relative energy loss as follows
\begin{equation}
\delta(\kappa) = {1 \over E_e N_e} \ \int E_\gamma \
dN_\gamma ={ N_0 \over   N_e}  \  \int_0^1 \Phi(y,\kappa) dy \ .
\label{12}
\end{equation}

Introducing the transformation 
\begin{equation}
x= {y\over 1-y} (1+z)
\label{12a}
\end{equation}
the integration over $z$ can be performed explicitly and the
energy loss is found in the form
\begin{equation}
\delta (\kappa)= {N_0\over   N_e} \int _0 ^\infty
f(x)\;{J(\omega )\over J(0)} \;dx,\;\;\; \omega= x {E_e \over 4
\gamma_e^2 \hbar}
\label{13}
\end{equation}
where
\begin{equation}
f(x)   =   {3(x^2-4x-12) \over 4 x^4} \ln{(x+1)} +{9 \over x^3} -
{3(x+2) \over 4(x+1)x^2} + {x+3 \over 8(x+1)^3} \ .
\label{14}
\end{equation}

The main contribution to the energy loss arises from the
integration region
\begin{equation}
{\omega \sigma_z\over c}\;=\;{x \over \kappa} \sim
{y\over \kappa (1-y)}  \stackrel{<}{\sim} 1 \
\label{16}
\end{equation}
where the ratio $J(\omega )/ J(0)$ (see (\ref{4a})) is of the order
of one.  Indeed, the region of small $x$ does not contribute
significantly to $\delta$ since the function $f(x)$ can be approximated
for $x \ll 1$ by
\begin{equation}
f(x)={1 \over 2 } - { 21 \over 20} \ x + \dots  \ .
\label{15}
\end{equation}
The region of large $x$ is suppressed due to the behaviour of
$J(\omega )/ J(0)$ and $f(x)$.

{}From the estimate (\ref{16}) it follows that    small photon
energies $E_\gamma \stackrel {<}{\sim} E_c$ dominate the energy
loss at  $E_c\ll E_e$, while at $E_c\gg E_e$ all energies
$E_\gamma$ are important up to the maximal value $E_\gamma
\approx E_e$.

For Gaussian beams Eq.~(\ref{13}) simplifies to
\begin{equation}
\delta(\kappa) = {N_0 \over N_e} \ G(\kappa) \ , \;\;\;
G(\kappa) = \int _0^\infty  f(x) \ \exp \left(-{x^2\over
\kappa^2} \right) \ dx \ .
\label{22}
\end{equation}
Its expansion at small and large values of parameter $\kappa$ is
found to be
\begin{equation}
\delta = \left\{
\begin{array}
     {r@{\quad  \quad}l}  \delta^{\rm class} \ \left( 1 -
  {21\over 10 \sqrt{\pi}} \ \kappa + \dots \right)
\ ,  & \kappa \ll 1\\
 {3 \over 4 } \ {
N_0 \over N_e} \ \left[ 1 - {\sqrt{ \pi} \over \kappa} \ \left(
\ln{  \kappa \over 2  } + {1 \over 6} - {\gamma_E  \over 2} \right)
\right]      \ ,  & \kappa \gg 1 \ .
\end{array}
\right.
\label{18}
\end{equation}
Here the energy loss in the classical limit is
\begin{equation}
\delta^{\rm class}   =  {\sqrt{\pi} \over 4} {N_0\over N_e} \  \kappa\ ,
\label{deltaclass}
\end{equation}
and $\gamma_E=0.5772$ denotes the Euler constant.

\begin{figure}[htb]
\begin{center}
\epsfig{file=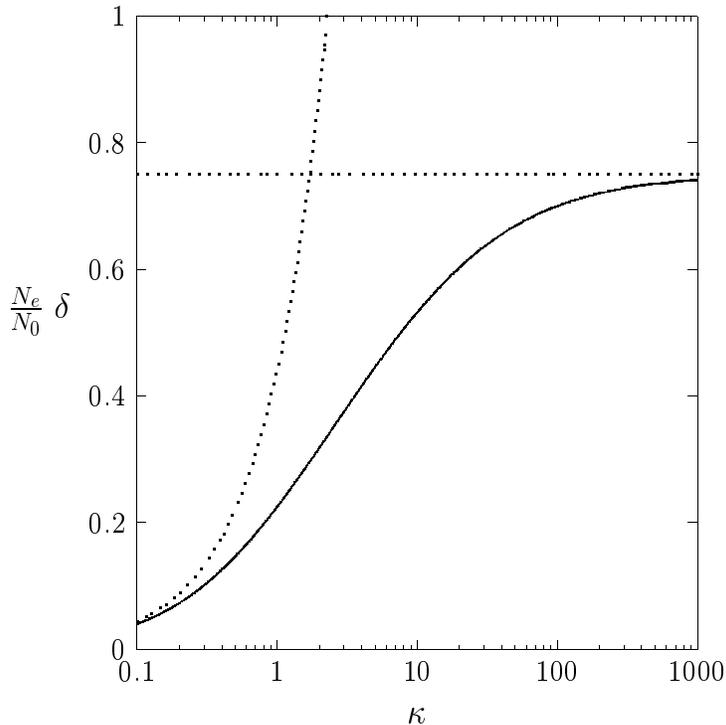,width=10cm}
\caption{Normalized energy loss $ (N_e / N_0) \delta$ as function
of the quantum parameter $\kappa$ (solid line), 
the classical and quantum limits are indicated by dotted lines
\label{qcbs-fig2}}
\end{center}
\end{figure}

The energy loss as function of the parameter $\kappa$ is
presented in Fig.~\ref{qcbs-fig2}
and in Table~\ref{qcbs-tab1}.
\begin{table}
\caption{\label{qcbs-tab1}
Few values of the function $G(\kappa)$ from Eq. (\ref{22}) }
\renewcommand{\arraystretch}{1.5}
\begin{center}
 \begin{tabular}{|c||c|c|c|c|c|c|c|c|c|c|c|}\hline
 $\kappa$
& 0.1 & 0.2 & 0.5 & 0.7 & 1.0 & 2.0 & 5.0 & 7.0 & 10.0 & 50.0 & 100.0  \\
\hline
$ G(\kappa)$
& 0.039 & 0.073 & 0.145 & 0.18 & 0.23 & 0.32 & 0.45 & 0.49 & 0.53 & 0.67
& 0.70 \\
\hline
\end{tabular}
\end{center}
\end{table}
For identical beams, the classical energy loss is well
approximated by
\begin{equation}
\delta^{\rm class} \approx 0.22  \ \alpha \ \eta^2 \ \kappa \, .
\end{equation}
We note that the energy loss for the coherent bremsstahlung
$\delta$ becomes constant at very large values of the quantum
parameter $\kappa$ (large beam energies).

It is interesting to compare the result derived for the energy
loss in CBS ($\eta \ll 1$) with that of the beamstrahlung ($\eta
\gg 1$). Since in the classical limit the energy loss is
determined by the square of the  electromagnetic fields, the
formula is the same for both cases (see Eq.~(\ref{deltaclass})).
The same argument holds if one studies the classical energy loss
as function of the impact parameter ${\bf R }$ between the two
colliding bunch axes. The whole effect is determined by the ${
\bf R } $ dependence of $N_0$ which has been described in detail
in \cite{ESS,Ginzyaf}, for flat  and round bunches.  Note
that an increase of the energy loss is expected for non head--on
collisions.

For beamstrahlung, the commonly used quantum parameter $\Upsilon$
is the ratio of the average critical energy for beamstrahlung $
\langle E_c^{BS}\rangle$ to the electron energy $E_e$
\cite{ChenR}. It can be expressed through the parameters $\kappa$
and $\eta$  (for identical beams)
\begin{equation}
\Upsilon = {2 \over 3} {<E_c^{BS}> \over E_e }= {5 \over 24} \ \eta
\ \kappa \ .
\label{220}
\end{equation}
An approximate expression  for the BS energy loss is given in
\cite{ChenR}
\begin{equation}
\delta ^{BS} = \delta^{\rm class} \left( 1 +
( 1.5   \Upsilon ) ^{2/3} \right)^{-2} \ .
 \label{bs }
\end{equation}
Comparing the loss in the extreme quantum cases we note the
different dependence on the incoming electron energy ($\kappa
\propto E_e$)
\begin{eqnarray}
{\delta \over \delta^{\rm class}} & \to & {  3  \over \sqrt{\pi}
\kappa} \ ,
\nonumber \\
{\delta^{BS} \over \delta^{\rm class}} & \to & {  0.58   \over
\Upsilon^{4/3}} \ .
\label{comp}
\end{eqnarray}

The parameter $\Upsilon$ is very small for most of the proposed
linear colliders. Therefore, the beamstrahlung is mainly
classical allowing to transform some known CBS properties of the
energy loss to colliders with large $\eta$. 

Let us consider, for
example, the TESLA collider \cite{TESLA} with flat transverse
beams $\sigma_x / \sigma_y = 600 \ \mbox{nm} / 6.5 \ \mbox{nm}
\approx 100$, $\sigma_z=0.5$ mm, the design beam energy $E_e =
500$ GeV with $N_e=N_p=1.8 \times 10^{10}$ particles per bunch.
The BS quantum parameter $\Upsilon $ is 0.053 leading to
$\delta=0.024$ which is about 30 \% smaller than the classical
energy loss. In this case the dependence of the energy loss on
the vertical beam displacement $R_y$ should be qualitatively the
same as that for the $DA \Phi NE$ collider   with approximately
the same transverse beam size ratio. For that collider the number
of produced photons and, therefore, the energy loss increases
almost two times at $R_y = 4 \sigma_y$ (see \cite{ESS}).

\section{Remark on the problem of beamstrahlung reduction using CBS
bunchlets}

A few years ago an idea has been proposed by Chen \cite{Chen87}
to reduce the BS energy loss. The discussion in \cite{Chen87} was
on a rough qualitative level only. In our terminology this
proposal can be formulated as follows:  partitioning a particle
bunch into a train of bunchlets, one could change the nature of
radiation from BS to CBS. This may not be a very
promising idea since such a proposal leads to a lot of
difficulties including an increase of the total bunch length.
If we omit such problems, we are able
to discuss this idea quantitatively.  In what follows we also
neglect, for simplicity, disruption effects.

We notice that beamstrahlung and CBS depend differently on
the number of bunchlets $n_b$. Denoting by the index $i$ the
contribution of an individual bunchlet $i$ we see from the above
formulae that the ratio $\delta_i^{BS} / \delta_i^{\rm class}$
depends on the parameter $\Upsilon_i \propto N_i/ \sigma_{zi}$
only.  Neglecting the longitudinal boundary effects,  $\Upsilon_i$
is the same as the quantum parameter $\Upsilon \propto N /
\sigma_{z }$ of the whole bunch. On the contrary, the parameters
$\eta$ and $\kappa$ of the whole bunch change to
\begin{equation}
\eta_i={ \eta \over n_b} \ , \ \ \ \kappa_i = n_b \ \kappa \ .
\label{etai}
\end{equation}
As long as the number of bunchlets is such that $\eta_i \gg 1$,
one remains in the beamstrahlung regime and practically no
energy reduction is obtained.

If $\eta_i$ is smaller than one, we arrive at the CBS regime
assuming that the spacing between two subsequent bunchlets is
such that their radiation is incoherent. In that case the ratio
of the energy losses $r_i=\delta_i^{CBS} / \delta_i^{class}$ depends   on
$\kappa_i$ and, therefore, on $n_b$.  The ratio  $r_i $ decreases with
increasing $n_b$ (remind that $r_i \to 2.3 /\kappa_i \propto
1/n_b$ at $\kappa_i \gg 1$). Therefore, one can in principle obtain an
considerable energy loss reduction for large enough number of
bunchlets.  To be precise, the following conditions for a
decreasing energy loss should be fulfilled.  First of all, the
parameter $\eta_i={ \eta / n_b}$ should be smaller than one,
hence, the number of bunchlets $n_b$ should be greater than
$\eta$. Additionally, to get a {\it considerable}
reduction one should use such values of
the parameter $\kappa_i = n_b \ \kappa \stackrel{>}{\sim} 1$
which leads to  $n_b \stackrel{>}{\sim} 1 /\kappa$. As a result,
we obtain the requirement

\begin{equation}
n_b \stackrel{>}{\sim} \max \left( {1\over \kappa} , \  \eta
\right) \ .
\label{nb}
\end{equation}

As a side remark we notice, that by choosing a relatively large
value of $n_b$ one could reach the situation where the length of
a bunchlet $\sigma_{zi} = \sigma_{z} / n_b$ can become smaller
than its transverse size $\sigma_{x}$.  In this case we are still
in the validity range of our formulae for CBS (where  we used the
restriction $\gamma_e \sigma_{zi} \gg \sigma_{x}$ only), though
it may be technically difficult to realize.

Let us give a numerical example to show how the discussed
reduction works.  
We consider the TESLA accelerator mentioned
above. For this collider we have $\eta = 85$ and $\kappa =
0.003$. The minimal number of bunches according to  (\ref{nb}) is
330.  For this value one obtains $\eta_i =0.26$ and $\kappa_i =1$
which leads to an reduction of the original energy loss by a
factor 1.4. Increasing the used number of bunches three times we
get an reduction of the energy loss by the factor 2.5 (notice
that in this case   $\sigma_{zi} = \sigma_{x}$).

\section{Coherent pair production}

\begin{figure}[htb]
\begin{center}
\epsfig{file=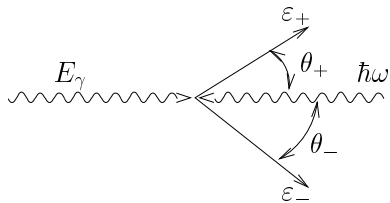,width=6cm}
\caption{  Kinematics for the process
$\gamma\gamma \rightarrow e^+e^-$, $\varepsilon_\mp$ denote the
energies and $\theta_\mp$ are the polar angles of the electron/positron
of the produced pair  \label{qcbs-fig3}}
\end{center}
\end{figure}

In this section we consider a cross-channel to CBS -- the
coherent pair production at $\gamma e$ colliders. This process
can be considered as $e^+e^-$ pair production in collisions of
initial photons of the energy $E_\gamma$ with the equivalent
photons of the energy $\hbar\omega$ corresponding to the
collective field of the electron bunch (see
Fig.~\ref{qcbs-fig3}).
The number of $e^+e^-$ pairs produced per single bunch crossing
is
\begin{equation}
dN_{e^+e^-} = dL_{\gamma\gamma}(\omega) \
d\sigma_{\gamma\gamma}(\omega,E_\gamma)\ ,
\end{equation}
where the spectral luminosity of the $\gamma\gamma$ collisions is
equal to (compare Eqs.~(39-41) of \cite{ESS})
\begin{eqnarray}
 dL_{\gamma\gamma}(\omega) &=& \frac{\alpha}{\pi}
        \frac{d\omega}{\omega} J(\omega) \ , \nonumber\\
  J(\omega)                &=& 4 \pi \int
         \frac{{\bf q}_\perp}       { {\bf q}_\perp^2}
         \frac{{\bf q}^\prime_\perp}{{{\bf q}^\prime_\perp}^2}
   F_e({\bf q}) F_e^* ({\bf q}^\prime)
   F_\gamma({\bf q}-{\bf q}^\prime)
\frac{d^2   q_\perp  d^2   q ^\prime_\perp} {(2 \pi)^4} \ .
\end{eqnarray}
Here $F_e$ and $F_\gamma$ are the form factors of the electron
and the initial photon bunches, respectively,
$d\sigma_{\gamma\gamma}$ is the cross section for the
$\gamma\gamma \rightarrow e^+e^-$ process.

Using the total pair production cross section as function of the
variable $v$ (the ratio of the invariant mass squared to the
threshold energy squared)
\begin{equation}
v= { \hbar \omega E_\gamma \over m_e^2 c^4 }
\end{equation}
we obtain the total number of produced pairs
\begin{equation}
N_{e^+e^-} = N_0 \int_1^\infty
\frac{\sigma_{\gamma\gamma}(v)}{\sigma_0}
 \; \frac{J(\omega)}{J(0)}\; {dv \over v} \ .
\end{equation}
The constant $N_0$ is defined in (\ref{1}), $\sigma_0= (8 \pi /
3) r_e^2$ is the Thompson cross section.  Choosing the main
spectral component (see (\ref{4a})) $\omega= c/ \sigma_z$ and $v
\approx 1$, one finds a characteristic energy $E_{\rm char}$ for
the produced lepton pair
\begin{equation}
E_{\rm char}= { m_e^2 c^3 \over \hbar} \sigma_z\ .
\end{equation}
For photon energies $E_\gamma > E_{\rm char}$ the pair production
becomes important.

\begin{figure}[htb]
\begin{center}
\epsfig{file=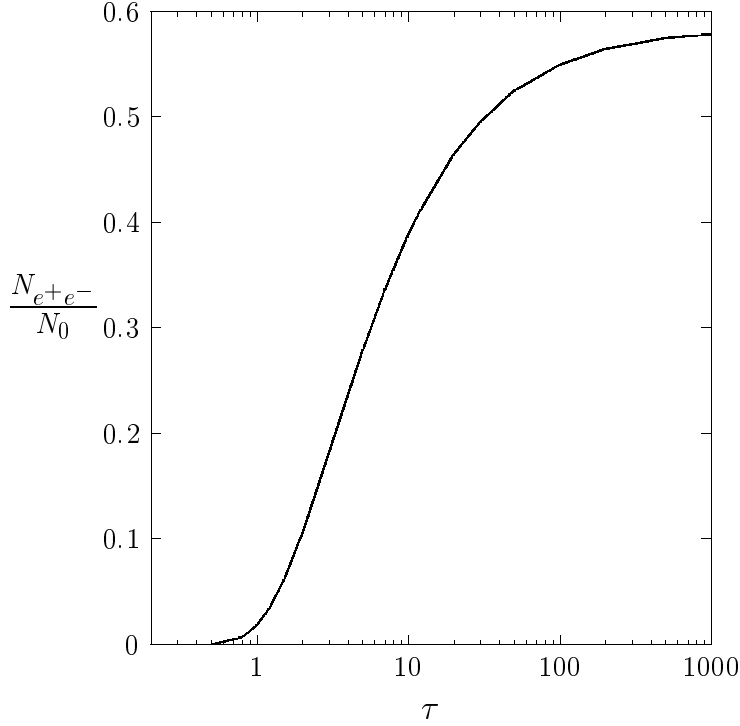,width=10cm}
\caption{
Normalized total number of produced pairs $N_{e^+e^-}/N_0$ as function of
the parameter $\tau$ (\protect\ref{tau})
 \label{qcbs-fig4}
}
\end{center}
\end{figure}

Introducing the ratio
\begin{equation}
\tau={E_\gamma / E_{\rm char}}
\label{tau}
\end{equation}
one gets for Gaussian beams ($\omega\sigma_z/c = v / \tau$,
$\sigma_z$ is the longitudinal electron bunch size)
\begin{equation}
N_{e^+e^-}(\tau)  =  N_0 \int_1^\infty \phi(v) \exp
\left(-{v^2\over \tau ^2}\right) dv
\end{equation}
with the function (comp. \cite{BLP} , \S 88)
\begin{eqnarray}
\phi(v) & =
&{\sigma_{\gamma\gamma}(v)\over \sigma_0} \ {1 \over v}
\nonumber
\\ & = & \frac{3}{8 v^4} \left[ 2 (v^2+v-\frac{1}{2}) \
\ln(\sqrt{v}+\sqrt{v-1}) - (v+1) \sqrt{v^2-v}\right]   \ .
\end{eqnarray}
It is not difficult to find that
\begin{equation}
N_{e^+e^-}(\tau) = N_0 \left\{\begin{array}{r@{\quad  \quad}l}
\frac{7}{12} \ ,  & \tau \gg 1\\  & \\
\frac{3} {64 }  \sqrt{2 \pi}\,  \tau^3\, \mbox{e}^{-1/\tau^2} \ , 
& \tau \ll 1 \ .
\end{array}\right.
\end{equation}
The normalized number of produced pairs as the function of the
ratio $\tau$ is shown in Fig.~\ref{qcbs-fig4}.  As expected a
rapid increase of the pair creation rate is observed for $\tau
\gg 1$.

Taking the differential cross section $d\sigma_{\gamma
\gamma}(\omega,E_\gamma,\varepsilon_\pm,{\bf p}_\pm)$ from
\cite{KSS} we obtain the energy--angular distribution of the
produced pairs ($\varepsilon_\pm$ and ${\bf p}_\pm$ are the
energy and momentum of the produced $e^\pm$, respectively)
\begin{equation}
dN_{e^+e^-}(\tau) = N_0 \frac{dz dx}{(1+z)^2}
\left[ \frac{3}{4} - \frac{3}{2}\, x(1-x)
\frac{1+z^2}{(1+z)^2}\right] \frac{J(\omega)}{J(0)}
\label{42}
\end{equation}
where
\begin{equation}
x = \frac{\varepsilon_+}{E_\gamma}\ , \ \ \
\varepsilon_- =   (1-x) E_\gamma \ , \ \ \
z = \frac{{\bf p}_{+ \perp}^2}{(m_e c)^2} \ , \ \ \
\hbar\omega = \frac{m_e^2c^4}{4 E_\gamma} \frac{1+z}{x (1-x)}.
\end{equation}
It follows from Eq.~(\ref{42}) that the main contribution to the
pair production is given by the region $z  \stackrel{<}{\sim} 1$,
i.e.\ $|{\bf p}_{+ \perp}| \stackrel{<}{\sim} m_e c$.

\begin{figure}[htb]
\begin{center}
\epsfig{file=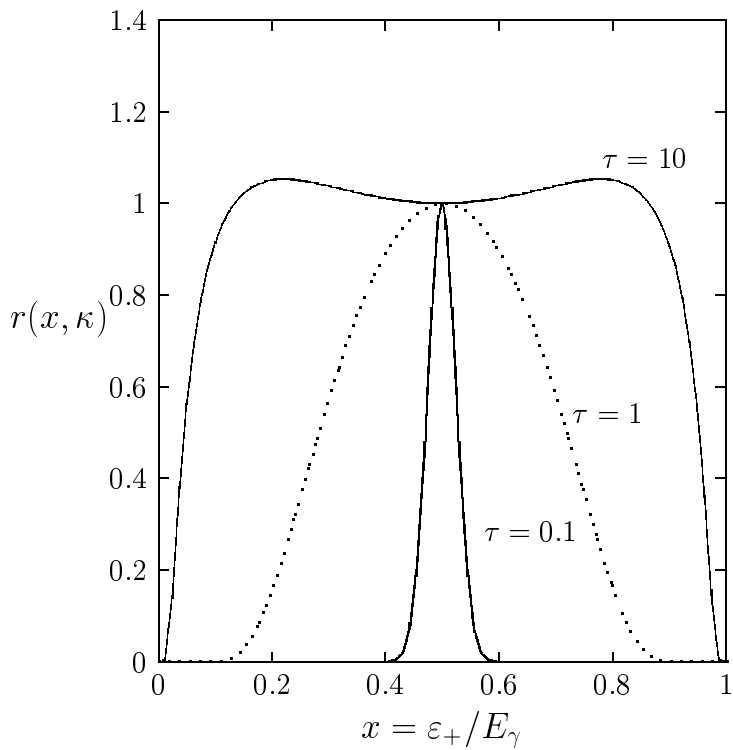,width=10cm}
\caption{Energy spectrum $r(x,\tau)= (d N _{e^+}(x,\tau)/dx)/
(d N _{e^+}(x,\tau)/dx)|_{x=0.5}$ for different $ \tau$ values
\label{qcbs-fig5}}
\end{center}
\end{figure}

Finally, integrating Eq.~(\ref{42}) over $z$, we obtain the
energy spectrum of produced $e^+$ for Gaussian beams
\begin{equation}
\frac{dN_{e^+}(x,\tau)}{dx} = N_0 \left[\frac{3}{4}
\Phi_2(u)-x(1-x)\Phi_1(u)\right], \hspace*{1cm}
u = \frac{1}{\tau} \ \frac{1}{4  x(1-x)}
\label{energysp}
\end{equation}
where the functions $\Phi_1(u)$ and $\Phi_2(u)$ are defined by
Eqs.~(\ref{13}) and (\ref{14}). The corresponding spectra for
different values of the parameter $\tau$ (normalized at the
energy fraction $x= 0.5$) are shown in Fig.~\ref{qcbs-fig5}.

For small values of $u$ ($u \ll 1$) the spectrum can be
approximated by
\begin{equation}
\frac{dN_{e^+}(x,\tau)}{dx} = N_0 \left[\frac{3}{4}- x(1-x)\right] \ .
 \end{equation}
In this limit, the energy spectrum becomes similar to the
well-known distribution of $e^+e^-$ pairs for the
photoproduction off nuclei (see \cite{BLP}, \S 94).  For large
values of $u$ ($u \gg 1$) the spectrum is suppressed like
\begin{equation}
\frac{dN_{e^+}(x,\tau)}{dx}=
N_0 \frac{3}{8} \left[ 1- 2 x (1-x)\right] \frac{1}{
 u^2} \exp (-u^2) \ .
\label{energysplim}
\end{equation}
For $\tau \ll 1$ it follows from
(\ref{energysp},\ref{energysplim}) that  the produced electrons
and positrons have approximately equal energies, $\varepsilon_+
\approx \varepsilon_-$. Additionally, taking into account ${\bf
p}_{+ \perp}=-{\bf p}_{- \perp}$, their polar angles are
approximately equal $\theta_+ \approx \theta_-$.

\section{Conclusions}

In the present paper we have calculated quantum effects of
coherent bremsstrahlung at colliders with short bunches.  The
calculation is based on the earlier developed equivalent photon
approximation for coherent processes.

The spectrum and the energy loss are found as function of the
quantum parameter $\kappa$ (\ref{01}). The classical regime
is given by $\kappa \ll 1$, the extreme quantum limit corresponds to 
$\kappa \gg 1$.  Quantum corrections to the classical limit as well as
corrections to the extreme quantum cases are found.  It was shown
that the electron energy loss changes its behaviour with the
growth of the parameter $\kappa \propto E_e$.  In the classical
case we have $\delta \propto \kappa$ while in the extreme quantum
regime $\delta $ becomes a constant.

As an example, the obtained formulae have been used to discuss
quantitatively the proposal \cite{Chen87} to reduce the
beamstrahlung energy loss using CBS bunchlets.  The quantum
effects in CBS may be also important for linear superconductive
colliders recently under discussion. For these colliders it is
planned to use the bunches of particles not only once but
repeatedly which leads to the requirement of small deflection
angles.

Finally, the coherent $e^+e^-$ pair production for $\gamma e$
colliders with short bunches is discussed.  The total number of
pairs is calculated, the energy spectrum and the energy--angular
distribution are presented.

\section*{Acknowledgments}
We are very grateful to V.~Balakin, R.~Brinkmann and D.~Schulte
for useful discussion. V.G.~Serbo acknowledges support of the
S\"achsisches Staatsministerium f\"ur Wissenschaft und Kunst, of
the Naturwissenschaftlich--Theoretisches Zentrum of the Leipzig
University and of the Russian Fond of Fundamental Research.
R.~Engel is supported by the Deutsche Forschungsgemeinschaft
under grant Schi 422/1-2.

\begin{appendix}

\section {Appendix: Approximation of constant $N_0$}

The dimensionless constant $N_0$ (\ref{1}) depends on the
transverse densities of the electron and positron bunches.  In
the case of Gaussian beams this quantity can be calculated in a
form of a one--dimensional integral \cite{Ginzyaf}.  For
identical beams the limits of round ($\sigma_x=\sigma_y$) and
flat ($\sigma_y \ll \sigma_x$) beams are known exactly
\cite{Ginzyaf,ESS}
\begin{equation}
N_0^{\rm round} = c^{\rm round} \ \alpha \ N_e \ \eta^2, \;\;\;
c^{\rm round}={ 16 \over 3 \pi} \ln{4 \over 3} \ \approx \  0.4884
\label{round}
\end{equation}
and
\begin{equation}
N_0^{\rm flat} =  c^{\rm flat} \ \alpha \ N_e \ \eta^2\ , \;\;\;
c^{\rm flat} ={ 8 \over 9 \sqrt{3}}   \ \approx \ 0.5132
\label{flat}
\end{equation}
where $\eta$ is given by (\ref{0}).
In the estimate (\ref{111}) we have assumed
\begin{equation}
c^{\rm round} \approx c^{\rm flat} \approx 0.5\ .
\end{equation}

\begin{figure}[htb]
\begin{center}
\epsfig{file=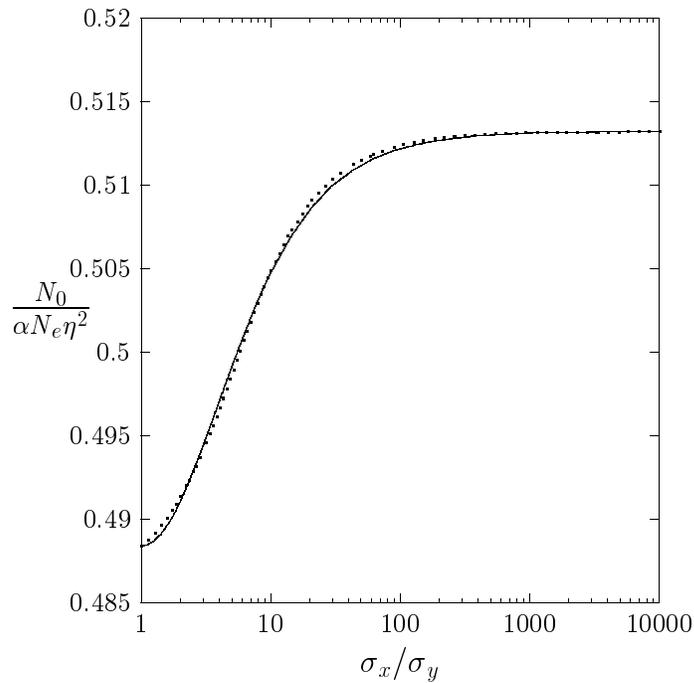,width=10cm}
\caption{Comparison of exact (solid line) and approximate (dotted line)
behaviour of $N_0/(\alpha N_e\eta^2)$ as function
of $\sigma_x / \sigma_y$ for identical beams
\label{qcbs-fig6}}
\end{center}
\end{figure}

A simple interpolation formula can be given parametrizing the
dependence of $N_0/(\alpha N_e\eta^2)$ for identical Gaussian
beams on the ratio $\sigma_x / \sigma_y$
(independent on the other bunch parameters) within 0.11 per cent
accuracy
\begin{equation}
{N_0 \over \alpha N_e \eta^2} =
c^{\rm round} + (c^{\rm flat}-c^{\rm round})
{ 2 \over \pi} \arctan\left[ 0.191 \left(  {\sigma_x
\over \sigma_y} -1 \right) \right] \ .
 \label{approx}
\end{equation}
In Fig.~\ref{qcbs-fig6} we show
the exact behaviour of $N_0/(\alpha N_e\eta^2)$  (full line) together
with the presented approximation (dotted line).

\end{appendix}

\end{document}